%% file: main.tex
\documentclass[sigconf,10pt]{acmart}

\usepackage{tikz}
\usepackage{amsmath}

\usepackage{filecontents}

\usepackage{xspace}
\usepackage{subfigure}
\usepackage{threeparttable}
\usepackage{xurl}
\usepackage{amsfonts}
\usepackage{multirow}
\usepackage{pifont}
\usepackage{algorithm}
\usepackage{algpseudocode}
\usepackage{booktabs}
\usepackage{hyperref}

\newcommand{\ie}{{\em i.e., \xspace}}
\newcommand{\etc}{{\em etc\xspace}}
\newcommand{\eg}{{\em e.g., \xspace}}

\newcommand{\First}{{\em First}}
\newcommand{\Second}{{\em Second}}

\newcommand{\system}{{Palantír}\xspace}
\newcommand{\para}[1]{\vspace{3pt}  \noindent {\bf #1} \hspace{3pt}}

\newcommand{\squishlist}{
\begin{list}{$\bullet$}{
  \setlength{\itemsep}{0pt}
  \setlength{\parsep}{3pt}
  \setlength{\topsep}{3pt}
  \setlength{\partopsep}{0pt}
  \setlength{\leftmargin}{3.5mm}
  \setlength{\labelwidth}{1em}
  \setlength{\labelsep}{0.5em}}}
\newcommand{\squishend}{\end{list}}

\renewcommand\footnotetextcopyrightpermission[1]{}
\setcopyright{none}
\usepackage{bbding}

\begin{document}

\date{}

\title{\system: Towards Efficient Super Resolution for Ultra-high-definition Live Streaming}

\newcommand{\tsc}[1]{\textsuperscript{#1}}
\author{Xinqi Jin\tsc{1\dag}, Zhui Zhu\tsc{2\dag}, Xikai Sun\tsc{2}, Fan Dang\tsc{3}, Jiangchuan Liu\tsc{4},\\Jingao Xu\tsc{5}, Kebin Liu\tsc{3}, Xinlei Chen\tsc{6,7,8}, and Yunhao Liu\tsc{2,3\Envelope}}
  \affiliation{\institution{\tsc{1}School of Software, Tsinghua University \tsc{2}Department of Automation, Tsinghua University}
  \country{}}
  \affiliation{\institution{\tsc{3}Global Innovation Exchange, Tsinghua University \tsc{4}School of Computing Science, Simon Fraser University}
  \country{}}
  \affiliation{\institution{\tsc{5}School of Computer Science, Carnegie Mellon University}
  \country{}}
  \affiliation{\institution{\tsc{6}Shenzhen International Graduate School, Tsinghua University}
  \country{}}
  \affiliation{\institution{\tsc{7}Pengcheng Laboratory, China \tsc{8}RISC-V International Open Source Laboratory, China}
  \country{}}
\email{{jinxq21, z-zhu22, sxk23}@mails.tsinghua.edu.cn, dangfan@tsinghua.edu.cn,}
\email{jcliu@sfu.ca, jingaox@andrew.cmu.edu, kebinliu2021@tsinghua.edu.cn,}
\email{chen.xinlei@sz.tsinghua.edu.cn, yunhao@tsinghua.edu.cn}
\thanks{\dag Co-primary authors.
\\\Envelope Corresponding author.}

\renewcommand{\shortauthors}{}
\renewcommand\footnotetextcopyrightpermission[1]{}
\pagestyle{plain}

\begin{abstract}
Neural enhancement through super-resolution (SR) deep neural networks (DNNs) opens up new possibilities for ultra-high-definition (UHD) live streaming over existing encoding and networking infrastructure.
Yet, the heavy SR DNN inference overhead leads to severe deployment challenges.
To reduce the overhead, existing systems propose to apply DNN-based SR only on carefully selected anchor frames while upscaling non-anchor frames via the lightweight reusing-based SR approach.
However, frame-level scheduling is coarse-grained and fails to deliver optimal efficiency.
In this work, we propose \system, the first neural-enhanced UHD live streaming system with fine-grained patch-level scheduling.
Two novel techniques are incorporated into \system to select the most beneficial anchor patches and support latency-sensitive UHD live streaming applications.
Firstly, under the guidance of our pioneering and theoretical analysis, \system constructs a directed acyclic graph (DAG) for lightweight yet accurate SR quality estimation under any possible anchor patch set.
Secondly, to further optimize the scheduling latency, \system improves parallelizability by refactoring the computation subprocedure of the estimation process into a sparse matrix-matrix multiplication operation.

The evaluation results suggest that \system incurs a negligible scheduling latency accounting for less than 5.7\% of the end-to-end latency requirement.
When compared to the naive method of applying DNN-based SR on all the frames, \system can \textbf{reduce the SR DNN inference overhead by 20 times (or 60 times) while preserving 54.0-82.6\% (or 32.8-64.0\%) of the quality gain}.
When compared to the state-of-the-art real-time frame-level scheduling strategy, \system can reduce the SR DNN inference overhead by 80.1\% at most (and 38.4\% on average) without sacrificing the video quality.
\end{abstract}

\maketitle

\input{intro-v7}
\input{background-v3}

\input{system-v2}

\input{method-estimation-v2}

\input{method-parallelism-v3}
\input{evaluation-v2}
\input{future-work}
\input{relatedwork}
\input{conclusion}

\appendix
\input{appendix-v3}

\newpage
\input{ref.bbl}

\bibliographystyle{ACM-Reference-Format}
\bibliography{ref}

\end{document}

%% file: intro-v7.tex
\section{Introduction}
\label{sec:intro}
UHD videos such as 4K and 8K videos are expected to form a huge market worth more than \$1 trillion in the following few years~\cite{4K-market}.
More and more UHD live-streaming applications are deployed to revolutionize many aspects of our society.
For example, the UHD live streaming of major sports events such as the Olympics~\cite{Paris, Beijing} creates unprecedentedly immersive experiences for audiences.
Besides, with the real-time UHD video from inspection and surveillance cameras,  human operators can have a precise understanding of the spot and remotely take immediate actions to prevent emergent accidents~\cite{beach, wind-turbine}.

However, the bitrates of the encoded UHD videos are significantly larger than videos of lower resolutions and pose great challenges to existing network infrastructure.
Specifically, the bitrates of 4K videos can be as large as 45Mbps~\cite{youtube-bitrate}, while the worldwide average uplink bandwidth of mobile broadband networks is only 11.07 Mbps~\cite{SpeedtestGlobalIndex}.
A common solution to this problem is using dedicated hardware encoders~\cite{kiloview-e3, aws-elemental-link, Paris}, which can encode the UHD video more efficiently and provide much lower bitrates.
However, hardware encoders are very expensive and typically cost hundreds to thousands of dollars.
Using fixed broadband networks for the uplink is an alternative solution, but it inevitably affects mobility and prohibits applications such as drone-based inspection~\cite{wind-turbine, remote-drone}.

Recently, neural enhancement has been proposed~\cite{nemo, neuroscaler, DcSR} and deployed~\cite{MSEDGE-vSR, JD-vSR, CommsEase-vSR, ZEGO-vSR} to improve video streaming.
It can potentially boost the broad deployment of UHD live streaming by allowing the streaming source to stream only a low-bitrate low-resolution (LR) video over the bandwidth-limited uplink and using a super-resolution (SR) deep neural network (DNN) to upscale the LR stream to its SR counterpart later.
However, as detailed in Sec.~\ref{sec:background-SR}, SR DNN inference incurs heavy computation overhead and deployment challenges.
Therefore, many research efforts have been aimed at optimizing the overhead.
NEMO~\cite{nemo} and NeuroScaler~\cite{neuroscaler} achieve this by categorizing frames into anchor frames and non-anchor frames: only anchor frames undergo computation-intensive DNN-based SR, while non-anchor frames are reconstructed via reusing the SR results of reference frames.
The most beneficial anchor frames are carefully selected so that a large quality gain is achieved with a small anchor frame set.
Nevertheless, these systems still fail to reduce the overhead optimally due to their coarse scheduling granularity and consequently insufficient utilization of videos' temporal redundancy.
For example, a beneficial anchor frame may still contain some existing objects that can be well reconstructed by reusing-based SR.

In this paper, we propose \system, the first patch-level neural-enhanced UHD live streaming system.
\system aims to further optimize the computation overhead via fine-grained scheduling, \ie selecting the appropriate type — anchor or non-anchor — for each patch (part of a single frame).
\system is centered around two primary goals.
The first is to accurately pinpoint the most beneficial anchor patches so that a smaller overhead can be achieved without sacrificing the quality gains.
The second is to minimize the scheduling latency to better support latency-sensitive applications such as public surveillance~\cite{GB/T-28181-2022, police}, interactive shows~\cite{youtube-mobile}, and remote drone operation~\cite{remote-drone, wind-turbine}.
As introduced next, \system integrates a theory-guided DAG-based quality estimation method and a parallelism scheme to meet the two goals.

\para{DAG-based quality estimation for patch-level scheduling (Sec.~\ref{sec:method-estimation}).}
Quality measurement for anchor selection violates the second goal due to the heavy measurement overhead.
Although some quality estimation strategies are proposed in NEMO and NeuroScaler for efficient selection, they are originally designed for other scenarios and can hardly be adapted to meet our two goals.
Tens of seconds are spent in NEMO for anchor frame selection in a video segment of a few seconds.
Extending the strategy to the patch level incurs even larger latency.
NeuroScaler uses a real-time yet inaccurate estimation method, and it mainly compensates for the inaccuracy by prioritizing certain frame types.
The underlying insight is that certain frame types have large degrees of reference.
Our preliminary experiments (detailed in \textbf{Sec.~\ref{sec:method-estimation-problems}}) validate the insight - the common language effect size~\cite{cles} reaches 97.4\% (close to the best case of 100\%) between prioritized frame types and normal frames.
However, the metric is as low as 58.9\% (close to the worst case of 50\%) between different sub-frame units of encoded videos, so type-based prioritization is unhelpful in patch-level scheduling.

In this paper, we propose to solve the estimation problem from a novel and theoretical perspective.
We systematically analyze the SR error (which is inversely related to the quality gains) accumulation process in neural enhancement for the first time (\textbf{Sec.~\ref{sec:method-estimation-analysis}}).
Incorporated with a few reasonable assumptions, the analysis suggests a lightweight DAG structure for quick and accurate simulation of the process.
The SR error incurred by any anchor patch set can be simulated based on the DAG.
Effective approximations are further proposed to easily construct the DAG without contradicting modern codecs and decoder softwares (\textbf{Sec.~\ref{sec:method-estimation-graph}}).

\para{Parallelized scheduling (Sec.~\ref{sec:method-parallelism}).}
We find that the DAG structure is highly irregular due to the complex reference relationship among patches, making it difficult to achieve the concurrent attainment of correctness and parallelism (\textbf{Sec.~\ref{sec:method-parallelism-complexity}}).
Yet, our measurement reveals that a sequential implementation of DAG-based estimation achieves the first goal but is still distant from the second goal.

We get out of the dilemma by utilizing the inherent characteristics of the DAG structure.
Specifically, we find that due to the acyclic reference relationship among frames, there exist no bi-directional edges between two groups of patches corresponding to two frames.
Therefore, we group patch nodes by their belonging frames and introduce both intra-set and inter-set parallelism to refactor the per-group computation process into a highly parallelizable sparse matrix-matrix multiplication (SpMM) operation.
Such optimization reduces the latency by more than 200 times without changing the scheduling result and simultaneously meets the two goals (\textbf{Sec.~\ref{sec:method-parallelism-solution}}).

The key contributions of the paper include:
\begin{itemize}
\item 
\system is the first to enable efficient SR enhancement for UHD live streaming via fine-grained scheduling.
\item 
\system can effectively identify the most beneficial anchor patches by DAG-based quality estimations.
The DAG modeling is based on our pioneering, systematic, and formalized analysis of SR error propagation.
\item 
By utilizing the inherent characteristics of the DAG structure, \system can be optimized via parallelism to further reduce the DAG-based anchor selection latency by more than 200 times while not affecting the selection results.
\item
We conduct extensive evaluations to demonstrate that \system significantly reduces the overhead of neural enhancement and incurs a negligible latency.
\end{itemize}

The rest of the paper is structured as follows.
Sec.~\ref{sec:background} reviews preliminary knowledge.
An overview of \system is outlined in Sec.~\ref{sec:system}.
We detail our theory-guided DAG-based estimation method in Sec.~\ref{sec:method-estimation} and the latency optimization strategy for estimation-based scheduling in Sec.~\ref{sec:method-parallelism}.
We evaluate \system in Sec.~\ref{sec:evaluation}.
A discussion on limitations and future work is conducted in Sec.~\ref{sec:future-work}.
Related work is reviewed in Sec.~\ref{sec:related-work} and the paper is concluded in Sec.~\ref{sec:conclusion}.

%% file: background-v3.tex
\section{Background}
\label{sec:background}

\subsection{Primer on SR Streaming}
\label{sec:background-SR}

SR DNNs typically incorporate convolution layers and modules specially designed for upscaling (\eg deconvolution~\cite{FSRCNN}, pixel shuffle~\cite{pixelShuffle}, \etc).
We refer readers to \cite{SRSurvey} for further details of the SR DNNs.

Currently, there are two common deployment models for SR enhancement.
One is to execute the SR DNN inference on the mobile streaming clients~\cite{MSEDGE-vSR, ZEGO-vSR, CommsEase-vSR, JD-vSR}, and the other is to execute the SR DNN on a cloud server so that a lot of audience for the same video can benefit from the one-time server-side enhancement~\cite{neuroscaler}.
Yet, SR DNNs are of high computation complexity and lead to severe deployment challenges in both deployment models.
In the first model, the battery of mobile clients can be easily drained.
Consequently, Microsoft Edge VSR disables its SR feature when the device is not being charged~\cite{MSEDGE-vSR}.
As for the second model, the heavy inference overhead appears in the form of the high monetary cost of using cloud servers (estimated to be at least \$1.690 per hour per 4K stream~\cite{neuroscaler}).
Lowering the overhead is essential to a broader application of SR enhancement.

\subsection{Reusing-based SR}
\label{sec:background-cache}

\begin{figure}[tbp]
     \centering
     \includegraphics[width=\linewidth]{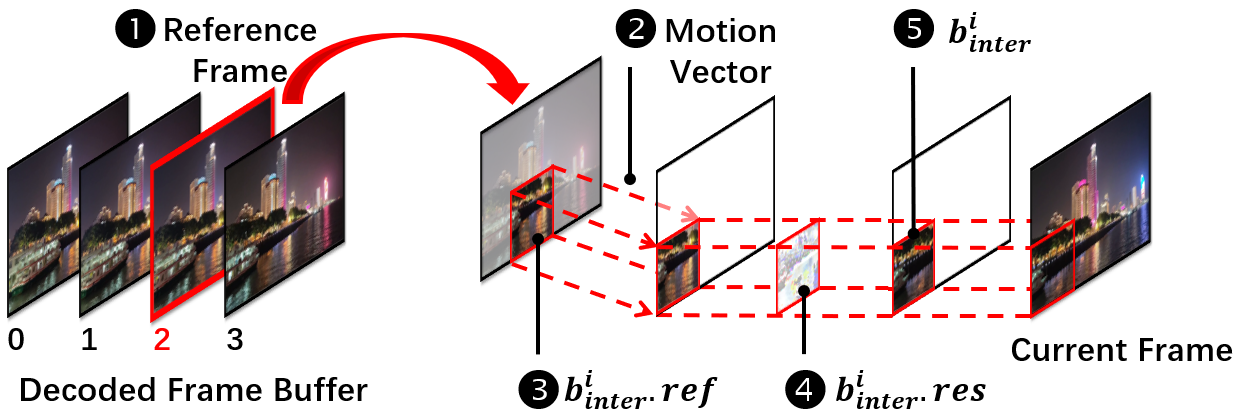}
    \caption{Video decoding pipeline.}
     \label{fig:decoding}
\end{figure}

To enable low-cost SR enhancement, researchers have built the NEMO~\cite{nemo} system, where an SR-enhanced decoder is adopted to reduce the cost by using video temporal redundancy.
In the SR decoder, video frames are categorically divided: anchor frames are upscaled via DNN-based SR, while non-anchor frames are quickly upscaled via reusing-based SR.
To appreciate the intricacies of reusing, a basic understanding of video codecs is essential.
Video coding predominantly encodes a block through inter coding; it identifies a similar reference block from a prior frame and only stores the subtle difference or residual between the two blocks.
A reference index is retained in the coded video to identify the frame containing the reference block or the reference frame, while a motion vector is stored to represent the potential spatial offset between the current and reference blocks (owing to object movements or camera shifts).
To decode an inter-coded block $b_{inter}^i$ (\ie \ding{206} in Fig.~\ref{fig:decoding}), the decoder first parses the reference index to determine its reference frame (\ding{202}) from the decoded frame buffer and then parses the motion vector (\ding{203}) to determine its reference block $b_{inter}^i.ref$ (\ding{204}). 
The reference block is added to the decoded residual $b_{inter}^i.res$ (\ding{205}).
This can be formally expressed as
\begin{equation}
\label{formula:inter-decoding}
    b_{inter}^i=b_{inter}^i.ref+b_{inter}^i.res.
\end{equation}
An alternative to inter-coding is intra-coding, which, while similar, leverages spatial redundancy and encodes an intra-coded block $b^i_{intra}$ by storing the intra-frame residual.
We refer the readers to the technical specifications~\cite{vp9-spec} for further details.
Note that patches are not the same as encoding blocks in this paper:
(1) blocks can have different sizes, while the sizes of all patches are the same to ease fine-grained scheduling;
(2) a block is either intra-coded or inter-coded, while a patch is either an anchor or non-anchor.
We leave selecting anchor patches of different shapes for future work.

\begin{figure}[tbp]
     \centering
     \includegraphics[width=\linewidth]{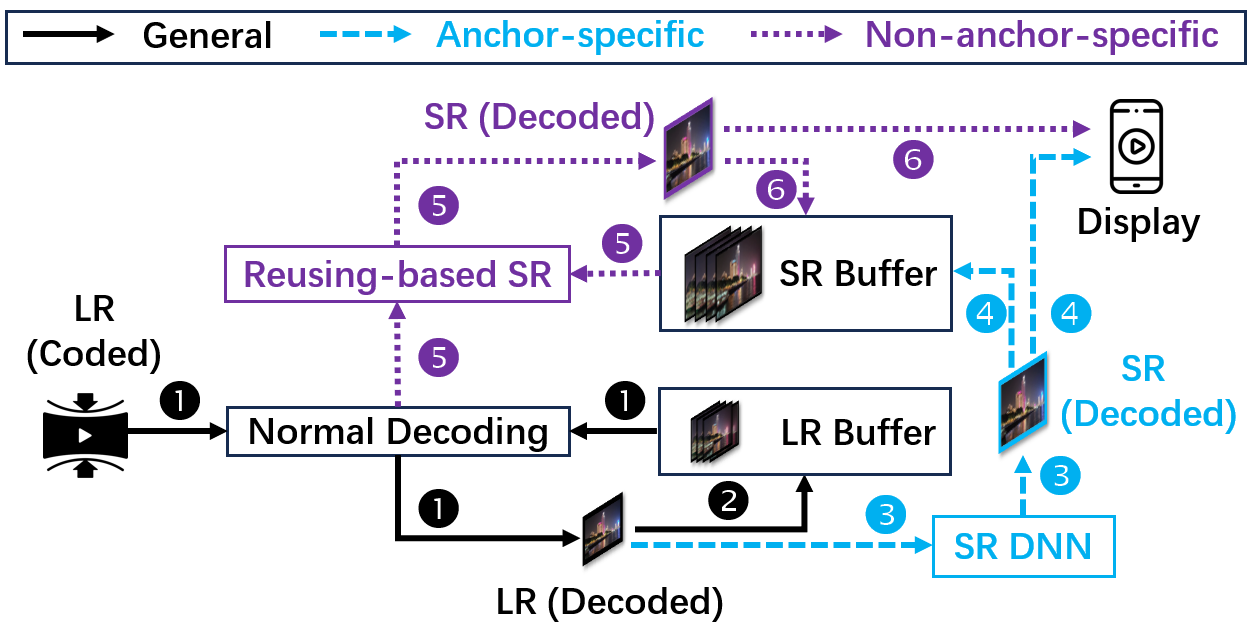}
    \caption{SR-integrated decoder overview.}
     \label{fig:sr-codec}
\end{figure}

The SR decoder in NEMO~\cite{nemo} is developed based on the open-source Google libvpx VP9 decoder~\cite{libvpx-repo}.
As shown in Fig.~\ref{fig:sr-codec}, the SR decoder first decodes a frame into its LR version by referring to decoded frames in the LR buffer (\ding{202}) and can insert it into the LR buffer for future frames (\ding{203}), just as a standard decoder.
Along with the LR video, a cache profile is also downloaded, each bit of which indicates whether a frame is an anchor or non-anchor.
If the current frame is an anchor, the LR version will be fed into the SR DNN to obtain the SR version (\ding{204}), which may be inserted into the SR buffer for future reuse (\ding{205}).
Otherwise, the SR version is obtained via reusing-based SR (\ding{206}).
The reusing-based SR uses a process similar to that in Fig.~\ref{fig:decoding} to decode every inter-coded block $b_{inter}^i$ to its SR version,  except that the motion vector is scaled (\eg by 4 times, when the LR and SR frames are 240p and 960p, respectively), the residual is upscaled by bilinear interpolation to match the resolution of the super-resoluted block ($b_{inter}^i.SR$), and the same reference index is used to fetch cached frames from the SR buffer rather than the LR buffer.
We can formulate the process as
\begin{equation}
\label{formula:reusing}
    b_{inter}^i.SR = b_{inter}^i.ref.SR + interp(b_{inter}^i.res, scale),
\end{equation}
where $scale$ is the ratio of the width of the SR video to that of the LR video.
We also rewrite Eq. (\ref{formula:inter-decoding}) as 
\begin{equation}
\label{formula:inter-decoding-rewritten}
    b_{inter}^i.LR = b_{inter}^i.ref.LR + b_{inter}^i.res
\end{equation}
to distinguish between the LR and SR version of the same block in the SR decoder.
As for any intra-coded block $b_{intra}^i$ in the non-anchor frame, it is upscaled by applying bilinear interpolation on its decoded LR version, \ie
\begin{equation}
\label{formula:intra-SR}
    b_{intra}^i.SR = interp(b_{intra}^i.LR, scale).
\end{equation}
After upscaling every block to their SR versions, the SR version of the non-anchor frame may be inserted into the SR buffer for future reuse (\ding{207}).
More details are available in \cite{nemo}.
Based on the open-source SR decoder in NEMO, we develop a fine-grained SR decoder, where a larger cache profile indicates the type (\ie anchor or non-anchor) of every patch, and, based on their types, patches are upscaled via either DNN-based or reusing-based SR.

\begin{figure}[tbp!]
\includegraphics[width=\linewidth]{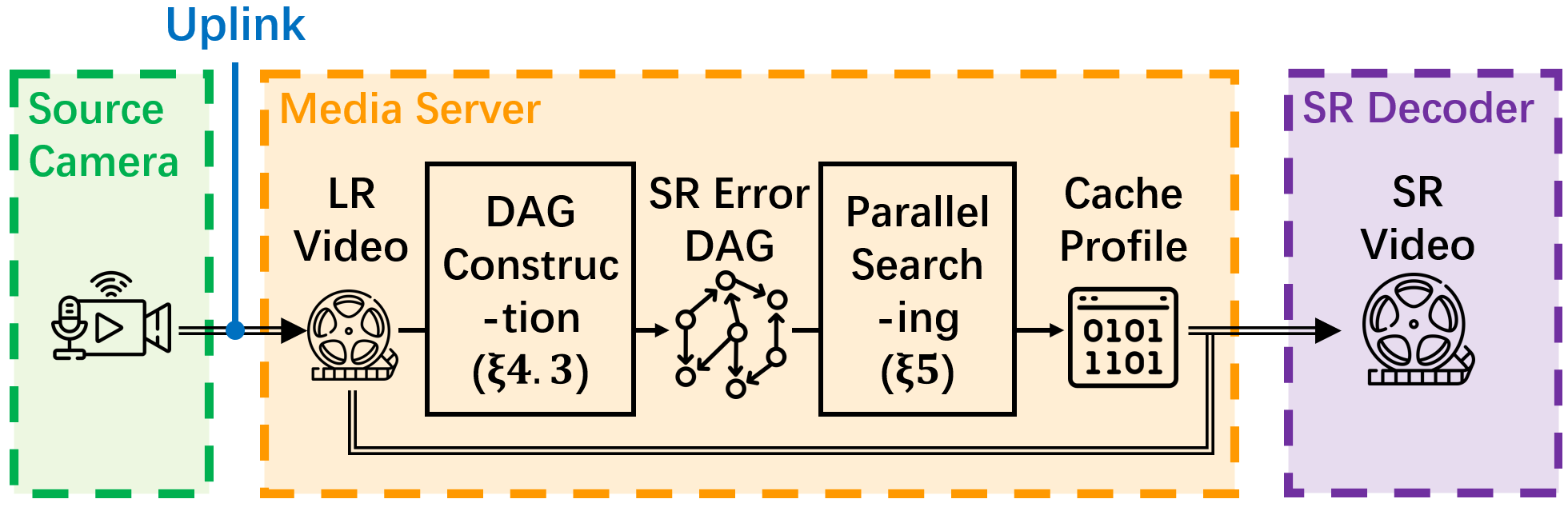}
\caption{\system overview.}
\label{fig:arch}
\end{figure}

\subsection{Estimation-based Anchor Selection}
\label{sec:background-anchor-selection}

Assuming that the DNN model is well suited for the video, selecting all patches as anchors naturally results in the best quality but also leads to the highest overhead.
Similarly, the number of anchor frames is limited in NEMO and NeuroScaler to reduce the cost.
These two systems thus greedily select the anchor frame that yields the highest quality until some budget or goal is met.
As measuring the video quality for greedy search is too time-consuming, NEMO and NeuroScaler approximate the video qualities with estimation values.
In NEMO, a heavy initial measurement phase (reported to take nearly one minute for a video segment of 4 seconds~\cite{nemo}) is required before conducting any estimation.
Extending the strategy to patch-level scheduling can even further slow down the measurement phase.
Since the strategy in NEMO fails to support live streaming due to its low anchor selection throughput, we do not use it as a baseline in our paper.
Alternatively, we choose to use the strategy introduced in NeuroScaler~\cite{neuroscaler} as our baseline.

%% file: system-v2.tex
\section{System Overview}
\label{sec:system}

\para{Scope.}
We aim to support UHD live streaming via SR enhancement.
As video frames are highly redundant, we propose to apply DNN-based SR only on carefully chosen anchor patches while upscaling non-anchor patches via the lightweight reusing-based SR mechanism.
This optimization is essential to improve the practicality of SR enhancement, considering the effect of SR inference on the battery life of mobile clients and the monetary costs of cloud-based SR inference.

\para{Design goals.}
\First, \system aims to pinpoint the most beneficial anchor patches, so that it can use a small anchor patch set to reduce the DNN inference cost while achieving a large quality gain.
\Second, \system should complete the fine-grained scheduling as quickly as possible, considering the stringent latency requirements in many video applications~\cite{GB/T-28181-2022, youtube-latency, remote-drone, aws-latency}.

\para{Workflow.}
The workflow of \system is shown in Fig.~\ref{fig:arch}.
Considering the limited uplink bandwidth~\cite{SpeedtestGlobalIndex} and the high bitrate of the original high-resolution (HR) video~\cite{youtube-bitrate}, the streamer only uploads the downsampled LR video to the media server.
The server generates an SR error DAG (\textbf{Sec.~\ref{sec:method-estimation-graph}}) for every LR video segment whose time duration equals the pre-defined scheduling interval.
The quality of the corresponding SR segment under any possible anchor patch set can be quickly estimated using the constructed DAG, and beneficial anchor patches are greedily searched via DAG-based quality estimation.
Two novel parallelism strategies are used to further accelerate the searching process without changing its results (\textbf{Sec.~\ref{sec:method-parallelism}}).
A cache profile is then created, every bit of which indicates whether a patch in the LR segment is an anchor or non-anchor.
In the existence of a powerful cloud server~\cite{neuroscaler}, both the LR segment and its corresponding cache profile are streamed to the server and then processed by the SR decoder on the server.
Or, alternatively, the SR decoder can be executed in the streaming client to perform cache profile-guided SR enhancement.

\para{Deployment Scenario.}
\system is designed for UHD live streaming but also supports videos of lower resolutions and video-on-demand services.
Our initial prototype is tailored for the VP9 codec, considering the engineering complexities involved in transitioning to other codecs.
However, the system is not restricted to VP9-specific functionalities and should be adaptable to a range of other codecs.

%% file: method-estimation-v2.tex
\section{DAG-based Quality Estimation}
\label{sec:method-estimation}

We first identify the problems with existing quality estimation methods.
To design a fine-grained and lightweight estimation method, we give a pioneering analysis of the SR error propagation process.
Based on our analysis, we propose DAG-based error propagation modeling and quality estimation.
We use reasonable approximations to easily determine the values of static attributes in the DAG.

\subsection{Problems with Existing Methods}
\label{sec:method-estimation-problems}

A natural way to estimate the video quality under a given anchor patch set is to extend the methods used in NEMO or NeuroScaler.
The strategy in NEMO is based on the observation that the quality gain of a frame is mostly determined by the most relevant anchor frame.
NEMO first enumerates every anchor frame set consisting of a single frame $f$ and measures the quality (denoted as $FQ(i|f)$) of every frame $i$ under every enumerated single anchor frame set $|f|$.
Then, NEMO estimates the quality under any anchor frame set AP as $FQ(i|AP)\approx max_{f\in AP}FQ(i|f)$.
Capitalized on the heuristics, NEMO requires a heavy measurement phase in nature and incurs a high latency (reported to be 59.6 seconds for a video segment of 4 seconds~\cite{nemo}) not suitable for live streaming.
The scheme can be easily extended to the patch-level granularity by firstly conducting measurements for all single anchor patch sets, but the latency of initial measurements can be further increased.

In contrast, the method in NeuroScaler~\cite{neuroscaler} is much more efficient.
It models the super-resolution error propagation process as a linked list, where each node corresponds to a frame and each pair of consecutive frames is linked.
The SR error of an anchor frame is assumed to be $0$, while the error of a non-anchor frame equals that of its preceding frame node plus the residual between the two frames.
An anchor set that leads to a lower sum of the errors over all frame nodes is regarded as leading to higher video quality.

However, such a modeling is too simplified and ignores the effect of the degree of reference.
In fact, each frame can have at most three reference frames in VP9~\cite{vp9-spec}, and its SR error directly depends on the error of every reference frame rather than that of a single preceding frame.
Conversely, some types of frames (\ie keyframes and alternative reference frames in the VP9 codec~\cite{vp9-spec}) may be referred to by many subsequent frames (rather than a single subsequent frame, as modeled by NeuroScaler).
To compensate for over-simplified modeling, NeuroScaler gives priorities to keyframes and alternative reference frames: 
a normal frame may be selected as the anchor only if all the keyframes and alternative reference frames (AltRefs) have been selected as anchors.
Yet, transferring the linked list-based modeling and the heuristic mitigation strategy to our context faces several challenges:

\begin{figure}[tbp]
     \centering
\includegraphics[width=0.75\linewidth]{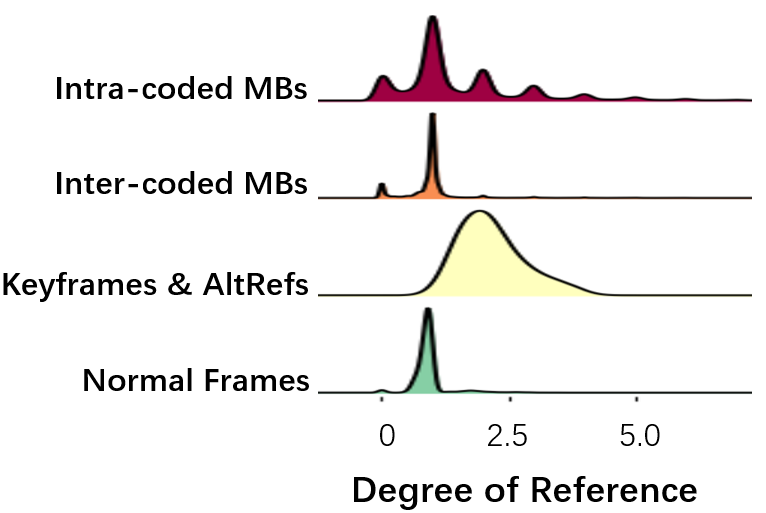}
    \caption{The distribution of degrees of reference for different types of frames and MBs.}
     \label{fig:preliminary-experiment}
  \vspace{-14pt}
\end{figure}

\squishlist
\item
\First, a patch can refer to a huge number of preceding patches, so it will be even more unreasonable if we model the SR error propagation process as multiple independent linked lists of patch nodes for fine-grained scheduling.
Denoting the number of rows in the patch grid as $nr$ and the number of columns in the patch grid as $nc$, and considering that each frame may refer to at most three frames in VP9~\cite{vp9-spec}, a patch in a VP9-coded video can refer to $3\cdot nr\cdot nc$ patches at most.
\item
\Second, we argue that the types of sub-frame encoding units such as macroblocks (MBs) do not implicitly imply their degrees of reference.
Here we conduct a preliminary experiment on the 480p version of the first benchmark video used in the evaluation part (Sec.~\ref{sec:evaluation}) to support our argument.
We separately measure the distribution of degrees of reference among different types of frames and macroblocks (including prioritized frame types, unprioritized normal frames, intra-coded MBs, and inter-coded MBs).
The degree of reference for a given frame (or MB) is quantitatively defined as $\frac{num\_references}{resolution}$, where $num\_references$ denotes the number of pixels that refer to the given frame (or MB) for inter coding and $resolution$ denotes the number of pixels in the given frame (or MB).
As shown in Fig.~\ref{fig:preliminary-experiment}, the degrees of references for keyframes and AltRefs are significantly greater than those of normal frames, while the distribution range of degrees of references for intra-coded MBs is similar to that for inter-coded MBs.
We further use the common language effect size (CLES)~\cite{cles} to quantize the results.
The CLES is defined as the probability that a value randomly sampled from one distribution will be greater than that from another distribution.
The CLES reaches 97.4\% (close to the best case of 100\%) between prioritized frame types and normal frames but is as low as 58.9\% (close to the worst case of 50\%) between different MB types.
Therefore, type-based prioritization is helpful in frame-level scheduling but hardly applies to patch-level scheduling.
\squishend

\subsection{Understanding SR Error Propagation}
\label{sec:method-estimation-analysis}

We next give an analysis of the SR error propagation process, which lays the foundation for our DAG-based quality estimation.
\squishlist
\item \textbf{Case \#1: non-anchor patches}.

\textbf{Analysis:}
every non-anchor patch $P$ may consist of multiple inter-coded blocks ($b_{inter}^1,...,b_{inter}^{n1}$) and intra-coded blocks ($b_{intra}^1,...,b_{intra}^{n2}$).
For analysis purposes, we split those blocks spanning multiple patches into multiple sub-blocks, each within a single patch.
The SR error of $P$ is:
\begin{align}
\label{formula:SR-error}
    &P.error\notag\\
    =&||P.SR-P.HR||_2^2\\
    =&\sum_{i=1}^{n1} ||b_{inter}^i.SR-b_{inter}^i.HR||_2^2 + \sum_{i=1}^{n2} ||b_{intra}^i.SR-b_{intra}^i.HR||_2^2\notag\\
    =&\sum_{i=1}^{n1} b_{inter}^i.error + \sum_{i=1}^{n2} b_{intra}^i.error.\notag
\end{align}

\begin{figure}[tbp]
     \centering
\includegraphics[width=\linewidth]{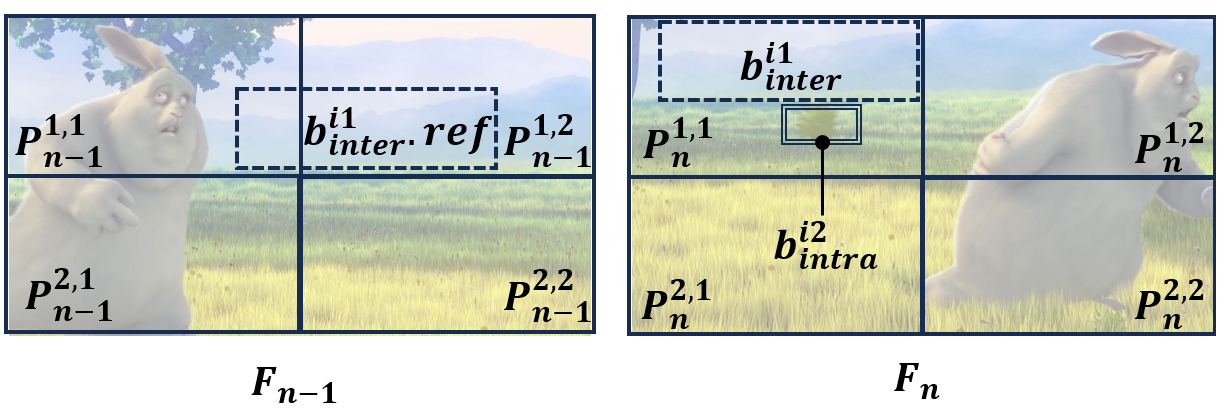}
    \caption{An example of SR error propagation.}
     \label{fig:patch-frame-block}
     
  \vspace{-14pt}
\end{figure}

\textbf{Example (see Fig.~\ref{fig:patch-frame-block}):}
$P_n^{1,1}.error$ equals the sum of $b_{inter}^{i1}.$ $error$, $b_{intra}^{i2}.error$, and the errors of many other blocks within $P_n^{1,1}$ (not marked due to the limited space).

\textbf{Conclusion \#1:
the SR error of a non-anchor patch is the sum of the SR errors of all inter-coded and intra-coded blocks within the patch.}

\textbf{Analysis:}
According to the equation (\ref{formula:reusing}), we can further write the SR error of an inter-coded block $b_{inter}^i$ as
\begin{align}
\label{formula:inter-error}
    &b_{inter}^i.error\\
    =&||b_{inter}^i.ref.SR+interp(b_{inter}^i.res, scale)-b_{inter}^i.HR||_2^2\notag\\
    \approx & b_{inter}^i.res.complexity + b_{inter}^i.ref.error,\notag
\end{align}
where 
\begin{align}
\label{formula:inter-res-complexity}
     & b_{inter}^i.res.complexity\\
    =&||interp(b_{inter}^i.res, scale)-(b_{inter}^i.HR-b_{inter}^i.ref.HR)||_2^2\notag\\
    =&||interp(b_{inter}^i.LR-b_{inter}^i.ref.LR, scale)-(b_{inter}^i.HR\notag\\
     &-b_{inter}^i.ref.HR)||_2^2\notag\\
    = &||interp(interp(b_{inter}^i.HR-b_{inter}^i.ref.HR, scale^{-1}), scale)\notag\\
     &-(b_{inter}^i.HR-b_{inter}^i.ref.HR)||_2^2\notag
\end{align}
and
\begin{align}
\label{formula:inter-ref}
     & b_{inter}^i.ref.error\\
    =& ||b_{inter}^i.ref.SR-b_{inter}^i.ref.HR||_2^2.\notag
\end{align}
Here, $b_{inter}^i.res.complexity$ relates to the \textbf{texture complexity} of the HR residual (\ie $b_{inter}^i.HR-b_{inter}^i.ref.HR$), since a HR residual with more complex texture details will experience a more significant deviation after the process of downscaling and re-upsampling, \ie $interp(interp(\cdot,$
$ scale^{-1}), scale)$.

Similarly, according to the equation (\ref{formula:intra-SR}), we have 
\begin{align}
\label{formula:intra-error}
    b_{intra}^i.error = b_{intra}^i.complexity,
\end{align}
where 
\begin{align}
\label{formula:intra-complexity}
    &b_{intra}^i.complexity\\
    =&||interp(b_{intra}^i.LR, scale)-b_{intra}^i.HR||_2^2\notag\\
    =&||interp(interp(b_{intra}^i.HR, scale^{-1}),scale)\notag\\
    &-b_{intra}^i.HR||_2^2\notag
\end{align}
relates to the \textbf{texture complexity} of the HR content (\ie $b^i_{intra}.HR$).

\textbf{Example (see Fig.~\ref{fig:patch-frame-block}):}
for $b^{i1}_{inter}$ in the frame $F_n$ (see Fig.~\ref{fig:patch-frame-block}), its SR error equals $b^{i1}_{inter}.res.complexity+b^{i1}_{inter}.ref.error$, where $b^{i1}_{inter}.ref$ is its reference block in $F_{n-1}$, a reference frame of $F_n$;
for $b^{i2}_{intra}$, its SR error equals $b^{i2}_{intra}.complexity$.

\textbf{Conclusion \#2:
the SR error of an inter-coded block depends on both the texture complexity of its HR residual and the SR error of its reference block; while the SR error of an intra-coded block depends on the texture complexity of its HR content.}

\textbf{Analysis:}
combining the equation (\ref{formula:SR-error}), (\ref{formula:inter-error}), and (\ref{formula:intra-error}), we have
\begin{align}
\label{formula:SR-error-2}
P.error    \approx P.TC+P.AE,
\end{align}
where 
\begin{align}
\label{formula:TC}
 P.TC=&\sum_{i=1}^{n1}b_{inter}^i.res.complexity \\
&+\sum_{i=1}^{n2}b_{intra}^i.complexity \notag
\end{align}
indicates the \textbf{texture complexity} of the HR content or the HR residual,
and
\begin{align}
\label{formula:AE}
 P.AE = \sum_{i=1}^{n1}b_{inter}^i.ref.error
\end{align}
is the \textbf{accumulated error} from depending blocks.

To reformulate the equation (\ref{formula:AE}) and simplify the SR error accumulation process along patches, we make the following assumption:

\textbf{Assumption \#1:
the per-pixel SR error within a patch is uniform - every pixel in the same patch shares exactly the same amount of error.
Or, formally speaking, we assume that 
\begin{align}
\label{formula:assumption}
p.error=||p.SR-p.HR||_2^2=\frac{P.error}{patch\_size}
\end{align}
holds for every pixel $p$ in some patch $P$.}

Denoting the set of reference patches of $P$ as $P^1$, ..., $P^{n3}$ and according to the \textbf{Assumption \#1}, we can reformulate the equation (\ref{formula:AE}) as
\begin{align}
\label{formula:AE-2}
    P.AE & = \sum_{i=1}^{n1}b_{inter}^i.ref.error \\
         & = \sum_{i=1}^{n1}\sum_{p\in b_{inter}^i.ref}p.error\notag\\
         & = \sum_{i=1}^{n3}W^i \cdot P^i.error,\notag
\end{align}
where the weight coefficient $W^i$ indicates the ratio of the number of referenced pixels in $P^i$ to the patch size.

\textbf{Example (see Fig.~\ref{fig:patch-frame-block}):}
for convenience, we assume that $b^{i1}_{inter}$ is the only inter-coded block in $P_n^{1,1}$.
The weight between $P_n^{1,1}$ and $P_{n-1}^{1,1}$ equals $0.105$, as the size of the intersecting region between $b^{i1}_{inter}.ref$ and $P_{n-1}^{1,1}$ is $0.105$ of the patch size; similarly, the weight between $P_n^{1,1}$ and $P_{n-1}^{1,2}$ equals $0.323$.
According to the equation (\ref{formula:SR-error-2}) and (\ref{formula:AE-2}), $P_n^{1,1}.error$ can be approximated as $P_{n}^{1,1}.TC + 0.105\cdot P_{n-1}^{1,1}.error+0.323\cdot P_{n-1}^{1,2}.error$.

From the equation (\ref{formula:SR-error-2}), (\ref{formula:TC}), and (\ref{formula:AE-2}) we can make the final conclusion:

\textbf{Conclusion \#3: under the Assumption \#1, the SR error of a non-anchor patch equals the weighted sum of the SR errors of its depending patches plus the texture complexities of its inter-coded HR residuals and its intra-coded HR contents.}

\item \textbf{Case \#2: anchor patches}.
Since it is common to train or fine-tune the SR DNNs to match the current video content in neural-enhanced streaming~\cite{neuroscaler, livenas, DcSR}, we make the following assumption:

\textbf{Assumption \#2: the SR errors of anchor patches are always $0$}.
\squishend

\subsection{DAG Construction}
\label{sec:method-estimation-graph}
According to the \textbf{Conclusion \#3}, the SR error of a non-anchor patch is the weighted sum of those of its depending patches plus its texture complexity.
A patch $P$ may depend on another patch $P^i$ for inter-coding only if the frame containing $P^i$ is a reference frame of the frame containing the patch $P$.
Since the reference relationship among frames is directed and acyclic, the SR error propagation process among patches is also directed and acyclic.
Therefore, we propose to use a directed acyclic graph (DAG) to represent the SR error origination and propagation process among patches, where every node corresponds to a patch and every edge indicates an inter-coding reference relationship.
A static \textit{weight} attribute is associated with every edge to reflect the degree of reference, corresponding to $W^i$ in the equation (\ref{formula:AE-2}).
Three attributes are associated with each patch node $P$:
\begin{itemize}
    \item The static $P.TC$ attribute represents the texture complexity (defined in the equation (\ref{formula:TC})) of $P$.
    \item The $P.is\_anchor$ attribute indicates whether the patch node is an anchor or non-anchor under a given anchor patch set.
    \item The \textit{P.error} attribute represents the SR error.
    When \textit{P.is\_anchor} equals $1$, \textit{P.error} equals $0$ (according to the \textbf{Assumption \#2}); otherwise, \textit{P.error} equals the weighted sum of the $error$ attributes of the predecessor nodes plus $P.TC$ (according to the \textbf{Conclusion \#3}).
\end{itemize}

\para{Problem formulation.}
We aim to set the appropriate values for the \textit{is\_anchor} attributes of all nodes to trade off the estimated quality and the inference overhead.
The quality is estimated as $-\sum_{P}P.error$.
The inference overhead is affected by the number of anchor patches (\ie $\sum_{P}P.is\_anchor$).

\para{Determining static attributes.}
At first glance of the last two lines of (\ref{formula:inter-res-complexity}) (or (\ref{formula:intra-complexity})), we need data only from the HR video to compute the block-level texture complexities and then aggregate the block-level results to obtain the static $P.TC$ attribute according to (\ref{formula:TC}).
Consequently, it seems that we only need to feed the HR video to the decoder and slightly modify the decoder to fulfill the computation of $P.TC$.
However, we identify two challenges of this method, with the former one applicable to all codecs and the latter one specific to codecs supporting invisible frames (\eg VP8, VP9, and AV1).
\First, the encoding process for the HR video and that for the LR video are independent.
Considering an inter-coded block $b^i_{inter}$ in the LR video and its corresponding block $b^i_{inter}.HR$ in the HR video, the two blocks may share different reference indices or irrelevant motion vectors due to the independent encoding processes.
Feeding the HR video to the decoder, the decoder can only obtain $b^i_{inter}.HR.ref$, the reference block of $b^i_{inter}.HR$ in the HR video, rather than $b^i_{inter}.ref.HR$, the block in the HR video that corresponds to $b^i_{inter}.ref$ in the LR video.
Yet, feeding both the LR video and the HR video to the decoder contradicts the conventional functionalities and structures of existing decoders.
\Second, codecs like VP8, VP9, and AV1 use invisible frames to achieve an effect similar to bi-directional prediction in H.26X codecs.
As the encoding processes for the LR and HR video are independent, an invisible frame may not have a corresponding invisible frame in the HR video.

\begin{figure}[tbp!]
\includegraphics[width=\linewidth]{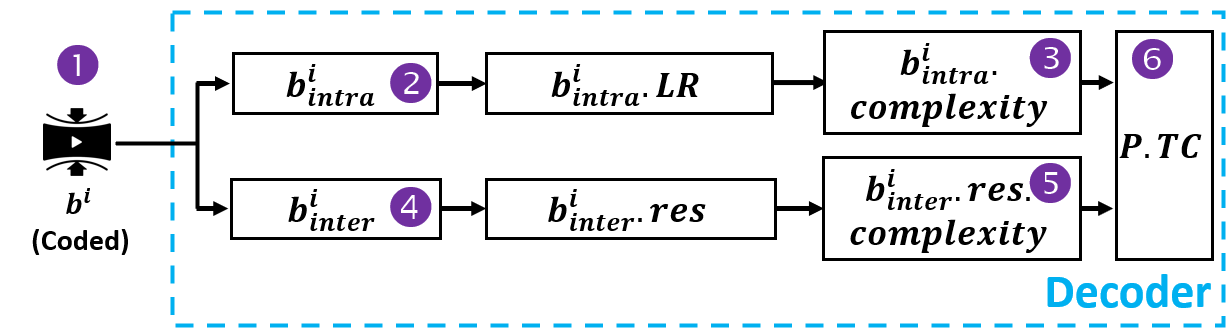}
\caption{Determining the value of the static \textit{TC} attribute.}
\label{fig:graph-construction-codec-palantir}

  \vspace{-14pt}
\end{figure}

To sidestep the two challenges, we propose to approximate the complexity of an inter-coded residual via
\begin{align}
\label{formula:inter-approx}
           & b_{inter}^i.res.complexity\\
        =  & ||interp(b_{inter}^i.res, scale)-(b_{inter}^i.HR-b_{inter}^i.ref.HR)||_2^2\notag\\
   \approx & ||interp(interp(b_{inter}^i.res, 0.5), 2)-b_{inter}^i.res||_2^2\notag
\end{align}
and approximate the complexity of an intra-coded block via
\begin{align}
\label{formula:intra-approx}
    & b_{intra}^i.complexity\\
        =  & ||interp(b_{intra}^i.LR, scale)-b_{intra}^i.HR||_2^2\notag\\
   \approx & ||interp(interp(b_{intra}^i.LR, 0.5), 2)-b_{intra}^i.LR||_2^2\notag
\end{align}

We base our approximation on the following observation:
if some content (\ie $b^i_{intra}.HR$) or some residual (\ie $b^i_{inter}.HR-b^i_{inter}.ref.HR$) is of high texture complexity, its downsampled version (\ie $b^i_{intra}.LR$ or $b^i_{inter}.res$) also tends to have high texture complexity.

With the above approximations, we can determine the \textit{P.TC} attribute by only resorting to data in the LR video.
We slightly modify the decoder to fulfill the computation process.
The workflow is shown in Fig.~\ref{fig:graph-construction-codec-palantir}.
While decoding a coded block $b^i$ (\ding{202}), it computes either the complexity of its content or its residual, based on its coding type, and then adds the value to $P.TC$ (\ding{207}), where $P$ is the patch containing the block.
In the case of intra-coding (\ding{203}), $b^i_{intra}.complexity$ (\ding{204}) is computed from the decoded $b^i_{intra}.LR$, following the equation (\ref{formula:intra-approx}).
In the case of inter-coding (\ding{205}), $b^i_{inter}.res.complexity$ (\ding{206}) is computed from the parsed $b^i_{inter}.res$, following the equation (\ref{formula:inter-approx}).

%% file: method-parallelism-v3.tex
\section{Parallel Searching}
\label{sec:method-parallelism}

\subsection{Performance Analysis}
\label{sec:method-parallelism-complexity}

We employ a greedy searching algorithm to iteratively select anchor patches based on the estimated quality.
We refer to the sequential implementation of this method as the vanilla \system.
Specifically, the estimation processes for different anchor patch sets are executed sequentially, and the \textit{error} attributes for different patch nodes under a given anchor patch set are also computed sequentially.
As demonstrated later in Sec.~\ref{sec:evaluation}, the vanilla \system meets the first goal but fails the second goal.

\begin{figure}[tbp]
\includegraphics[width=\linewidth]{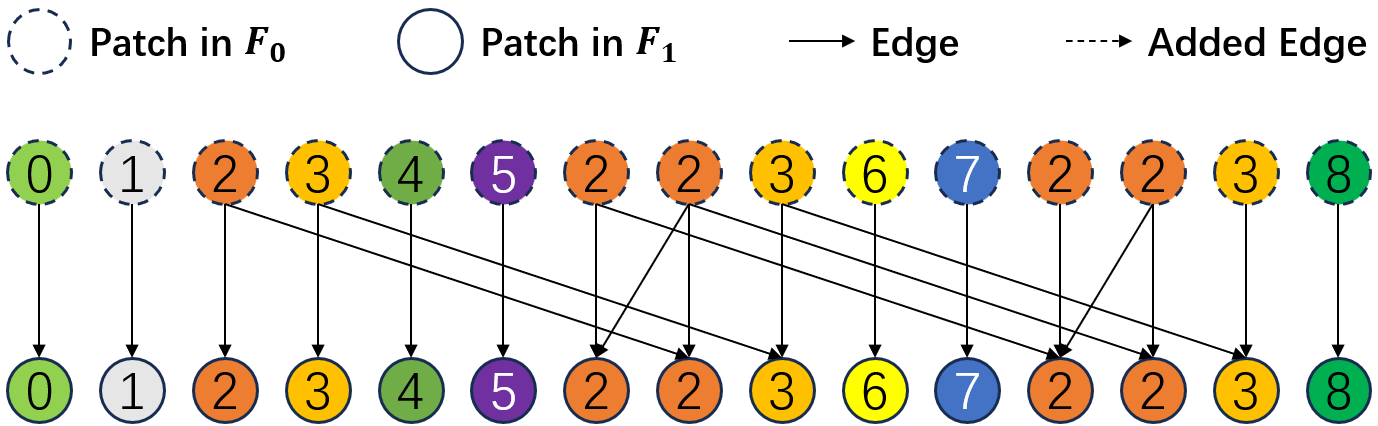}
\caption{A case study of quality estimation, with $nr\cdot nc=15$ and $F_0$ being a reference frame of $F_1$.
Sub-graphs are marked with different numbers and colors.}
\label{fig:case-study-original}
  \vspace{-14pt}
\end{figure}

To improve the scheduling latency, an intuitive method is to introduce parallelism into the DAG-based estimation process.
However, the complex reference relationship among patches leads to a highly irregular DAG structure, making it challenging to simultaneously guarantee correctness, parallelism, and data locality of computation.
We conduct a case study to demonstrate this.
For simplicity, we use only the first two frames of an LR segment to construct an SR error DAG (shown in Fig.~\ref{fig:case-study-original}) and identify two challenges.
\squishlist
\item
\First, to ensure the correctness of quality estimation, the \textit{error} attribute of any non-anchor node can be computed only after the \textit{error} attributes of all its predecessor nodes are computed.
The process can be parallelized by processing different nodes in different threads, only if there exists no directed path from one node processed in some thread to another node processed in another thread; otherwise, the correctness requirement may be violated.
As shown in Fig.~\ref{fig:case-study-original}, we can divide the DAG into nine disconnected sub-graphs and process different sub-graphs in different threads.
However, each sub-graph is rather small due to the sparse connections among patches, making it challenging to effectively utilize the SIMD features of modern CPUs.

\item 
\Second, while configuring data used by an individual thread to ensure memory locality is fairly straightforward, the scenario becomes more complex when multiple threads access data from various sub-graphs.
It requires an intricate thread-synchronization mechanism to maintain memory locality throughout the execution of multiple threads.
\squishend

\subsection{Parallelism Solution}
\label{sec:method-parallelism-solution}

We propose a novel strategy to enable parallel searching in \system.
The parallelized \system can generate the same anchor patch set as the vanilla \system as its parallelism mechanism does not hurt the correctness of the computation.
As demonstrated later in Sec.~\ref{sec:evaluation-latency} and~\ref{sec:evaluation-ablation}, the parallelized \system improves the DAG-based selection latency by more than 200 times and meets the two design goals simultaneously.

For clarity, we continue with the above case study in Sec.~\ref{sec:method-parallelism-complexity} to introduce our parallelism mechanism.
As shown in Fig.~\ref{fig:case-study-original}, edges always start from some patch in $F_0$ and point to some patch in $F_1$.
There exist neither edges along the opposite direction nor edges connecting patches within the same frame.
We will utilize this phenomenon to optimize \system via both intra-set and inter-set parallelism.
Note that the observed phenomenon is not accidental: edges indicate inter-frame coding references, and the coding reference relationship among frames is directed and acyclic in most codecs.
Therefore, our optimization should be applicable to many codecs.

\para{Intra-set parallelism.}
Based on the above phenomenon, we can follow the frame decoding order to enumerate the DAG for quality estimation, \ie firstly compute the \textit{error} attributes for the nodes in $F_0$ and then deal with the nodes in $F_1$.
The \textit{TC} attributes of nodes in $F_0$  (or $F_1$) can be denoted as a vector $TC_0$ (or $TC_1$).
Similarly, we use the notation $Error_i$  for the \textit{error} attributes and $Is\_anchor_i$ for the \textit{is\_anchor} attributes ($i=0,1$).
The \textit{weight} attributes of the edges can be denoted by a sparse $Weight$ matrix.
Note that the $Weight$ matrix is sparse since each patch in $F_1$ only refers to a limited set of patches in $F_0$ due to the temporal locality of videos.
The computation process can be formalized as
$Error_0 = TC_0 \circ Is\_anchor_0$ and $Error_1 = (TC_1 + Weight\cdot Error_0) \circ Is\_anchor_1$, where $\circ$ indicates the element-wise multiplication and $Weight\cdot Error_0$ is a parallelizable sparse matrix-vector multiplication (SpMV) operation.
Although the concurrent attainment of correctness, parallelism, and data locality for SpMV operations are also challenging, SpMV itself is a common operation in many application files and thus attracts many research and engineering efforts~\cite{SpMV1, SpMV2, SpMV3}.
Consequently, parallelized SpMV can be simply achieved by using a mainstream matrix-related computation package such as PyTorch.

\para{Inter-set parallelism.}
Batching is widely used to improve DNN inference throughput~\cite{batching} due to the effect of the data dimension on parallelism opportunities~\cite{TensorFlow, onthefly-batching}.
Therefore, we execute the quality estimation under several searched anchor sets in parallel by adding a batch dimension to both $Error_i$ and $Is\_anchor_i$.
The same $Weight$ matrix and $TC_i$ vectors are shared among different samples in the batch.
With the inter-set parallelism, the SpMV operations are converted into the sparse matrix-matrix (SpMM) operations, which are also well studied and supported for parallel implementation.

%% file: evaluation-v2.tex
\section{Evaluation}
\label{sec:evaluation}

We evaluate \system by answering three questions:
\squishlist
\item Does \system achieve the first design goal of selecting a beneficial anchor patch set and improving the efficiency of neural-enhanced UHD live streaming?
\item Does \system achieve the second design goal of incurring a negligible latency overhead for UHD live streaming?
\item How does each component of \system contribute to its overall performance?
\squishend

\subsection{Experimental Setup}
\label{sec:evaluation-setup}

\para{Implementation.}
We develop our decoder based on the open-source SR decoder in NEMO~\cite{nemo}.
We incorporate two novel modes into the decoder.
The first is to obtain the data required for graph construction (as introduced in Sec.~\ref{sec:method-estimation-graph}).
The second is to take both an LR video and a corresponding cache profile as input, and then upscale patches by either SR DNNs or reusing-based SR based on the cache profile.
\textbf{
The source code is available at \url{https://palantir-sr.github.io}.
}

\para{Hardware.}
We use a server with a 16-core AMD Ryzen processor as our media server, where graph construction and anchor selection are performed.
The scheduling latency is measured on the server.
We use a Xiaomi 12S smartphone, which was announced in July 2022 and equipped with the Qualcomm Snapdragon 8+ Gen 1 Mobile Platform, to measure the energy efficiency when running SR DNN inference on mobile receiver devices.

\para{Video.}
We download six popular 4k@30fps videos from YouTube.
To demonstrate the universality of \system, the videos contain six distinct categories, including \texttt{makeup review}, \texttt{computer gaming}, \texttt{skit}, \texttt{shopping}, \texttt{car review}, and \texttt{unboxing}.
We use FFmpeg (v3.4)~\cite{ffmpeg} to transcode the HR video into the 480p ($854\times 480$) LR version in real time.
We follow encoding guidelines to set the bitrate to \texttt{1800 kbps}, the encoding speed to \texttt{5}~\cite{google-guideline}, and the group of pictures (GoP) to \texttt{60} frames (\ie 2 seconds)~\cite{gop}.
We use the \texttt{-auto-alt-ref} option in FFmpeg to enable the alternative reference frame feature required by the anchor selection algorithm in NeuroScaler.
Unless noted otherwise, We use the first five minutes of each video in our evaluation.

\para{SR DNN.}
We adopt the DNN model of NAS~\cite{nas}.
We empirically set the number of residual blocks to 8 and the number of filters to 48.
The DNN upscales the resolution of the LR video by 4 times.
As the feasibility of online training for live streaming has been demonstrated~\cite{livenas}, we train the DNN model for each benchmark video.
When comparing the performance of different methods on the same video, the same DNN model is used for fairness.

\para{Anchor patch size.}
We use a patch size of $170\times 160$ to compensate for energy efficiency and latency.
Consequently, each LR frame consists of 15 patches.

\para{Baselines.}
We use three baselines in this part.
The first is the Per-frame baseline, which applies DNN-based SR on all the frames.
The second is the NeuroScaler baseline, which uses the algorithm in NeuroScaler~\cite{neuroscaler} to select the anchor frame set.
The third is the Key+Uniform baseline, which selects all the patches in the keyframe and equally spaced patches in the remaining frames as anchor patches.

\para{Scheduling interval.}
Unless otherwise noted, we use a scheduling interval equal to the GoP (\ie 2 seconds).

\para{Parallelism.}
The parallelized \system and the vanilla \system are two different implementations of the same selection method and lead to the same anchor patch set, so the results in Sec.~\ref{sec:evaluation-effectiveness} apply to both implementations.
The latency results in Sec.~\ref{sec:evaluation-latency} are obtained using the parallelized \system.
Finally, the two implementations are compared in Sec.~\ref{sec:evaluation-ablation}.

\begin{figure}[t]
\setlength{\abovecaptionskip}{-2pt}
\subfigure[\normalsize Setting: $m = 1$.]{\includegraphics[width=0.95\linewidth]{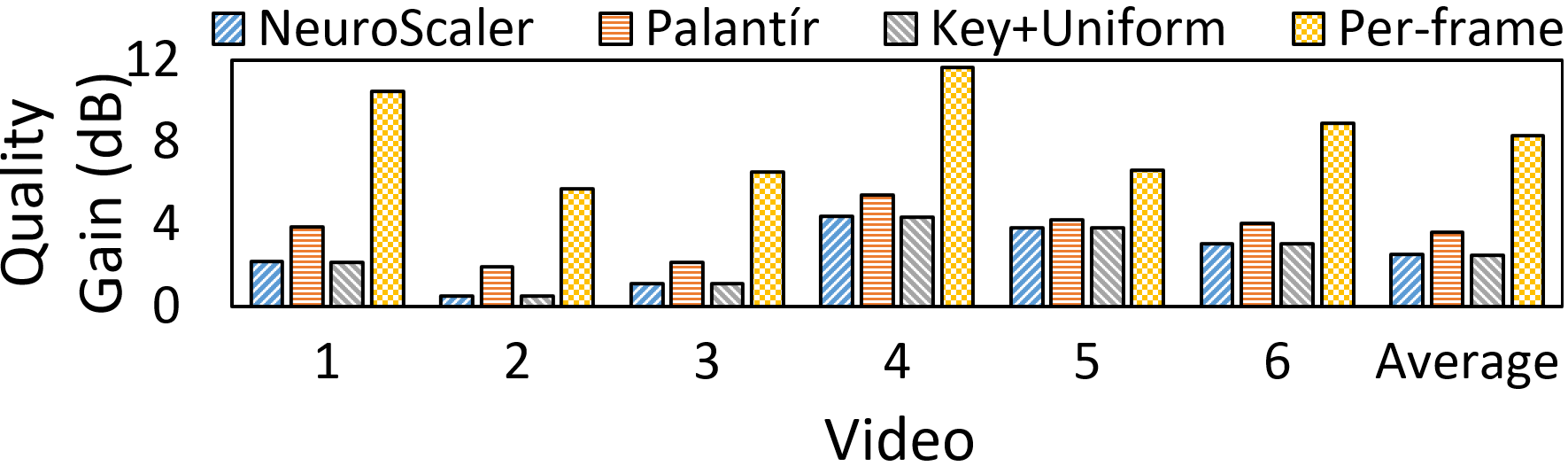}\label{fig:anchor-effectiveness-1}}
\subfigure[\normalsize Setting: $m = 2$.]{\includegraphics[width=0.95\linewidth]{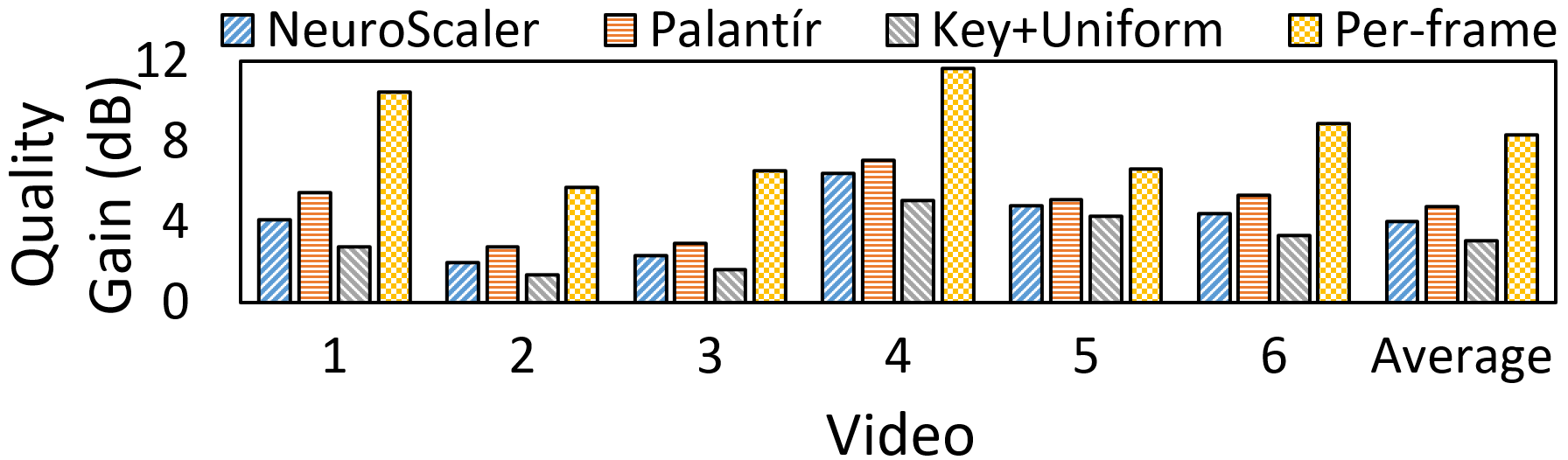}\label{fig:anchor-effectiveness-2}}
\subfigure[\normalsize Setting: $m = 3$.]{\includegraphics[width=0.95\linewidth]{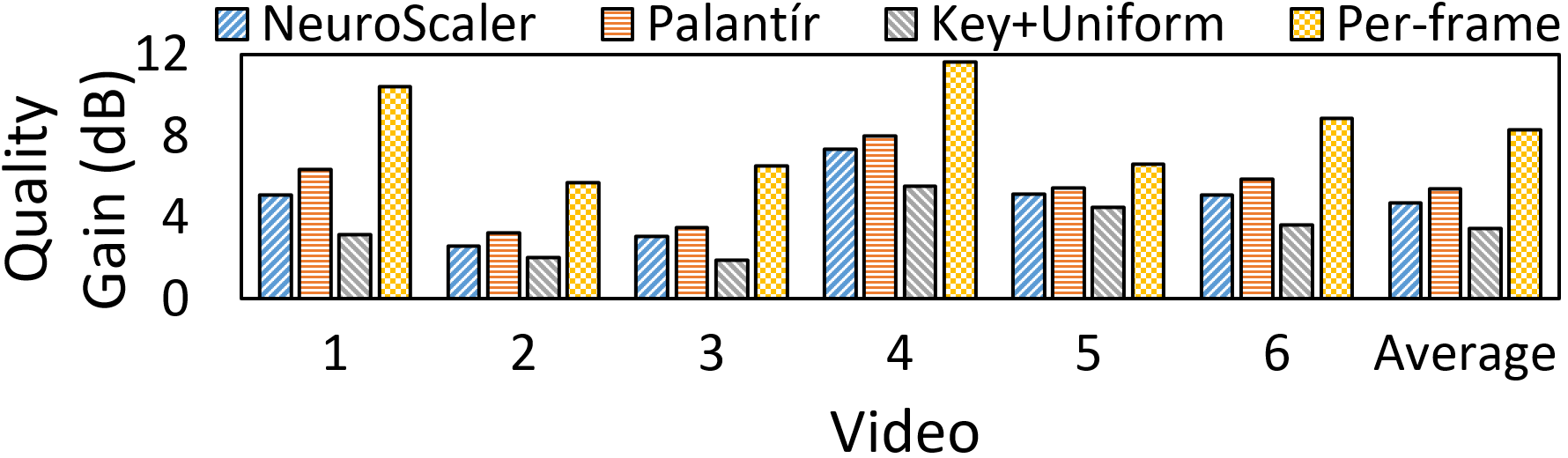}\label{fig:anchor-effectiveness-3}}
\caption{A comparison of anchor effectiveness.}
\label{fig:anchor-effectiveness}

  \vspace{-14pt}
\end{figure}

\subsection{Anchor Effectiveness}
\label{sec:evaluation-effectiveness}

\para{Quality Gain.}
We compute the peak-signal-to-noise-ration (PSNR) between the SR video and the original HR video to quantify the effectiveness of an anchor set.
To make a fair comparison, we keep the total sizes of the anchor regions the same, \ie compare the quality gain under the $m$-ary anchor frame set selected by the NeuroScaler baseline with that under the $(15\cdot m)$-ary anchor patch set selected by \system or the Key+Uniform baseline.
The only exception here is the Per-frame baseline, where all the frames are always treated as anchors and the size of the anchor regions is thus always larger than other methods.
Furthermore, we empirically limit $m$ to not be greater than $3$ since:
(1) $m=3$ can deliver quality gains that are comparable to the setting of applying DNN-based SR on all frames;
(2) further increasing the value of $m$ leads to a limited benefit yet a significant overhead.

The results are shown in Fig.~\ref{fig:anchor-effectiveness}, from which we have three observations:
(1) \system consistently outperforms the NeuroScaler baseline and the Key+Uniform baseline with its ability to identify beneficial patches.
\system boosts the quality gain of neural enhancement by 3.7 times at most and 1.4 times on average than the NeuroScaler baseline, or by 3.7 times at most and 1.7 times on average than the Key+Uniform baseline.
(2) The fine-grained scheduling-based Key+Uniform baseline even falls behind the coarse-grained scheduling-based NeuroScaler baseline, so the effect of fine-grained scheduling heavily depends on the anchor selection method.
(3) \system reduces the SR DNN inference overhead by 20 times with $m=3$ (or 60 times with $m=1$) while compared to the Per-frame baseline.
With such a remarkable overhead reduction, \system still preserves 54.0-82.6\% (or 32.8-64.0\%) of the quality gain of the Per-frame baseline.

\para{Energy Efficiency.}
We now examine how the anchor efficiency of \system transfers to energy efficiency when running the SR decoder on mobile devices.
We enable the developer mode on the Xiaomi 12S smartphone and record the average current over a specified period using the built-in power monitor software.
The detailed energy consumption measurement method is presented in \S~\ref{sec:appendix-energy}.

\begin{figure}[t]
\setlength{\abovecaptionskip}{-2pt}
\subfigure[\normalsize Setting: 1 anchor frame / 15 anchor patches.]{\includegraphics[width=0.9\linewidth]{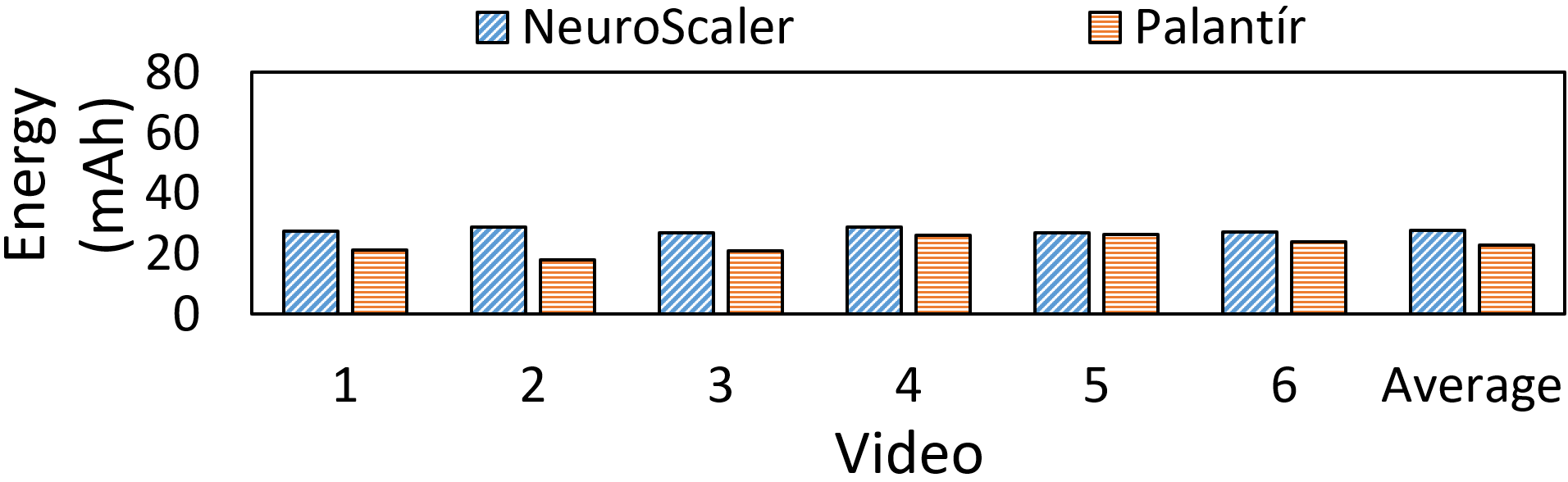}\label{fig:energy-effectiveness-1}}
\subfigure[\normalsize Setting: 2 anchor frames / 30 anchor patches.]{\includegraphics[width=0.9\linewidth]{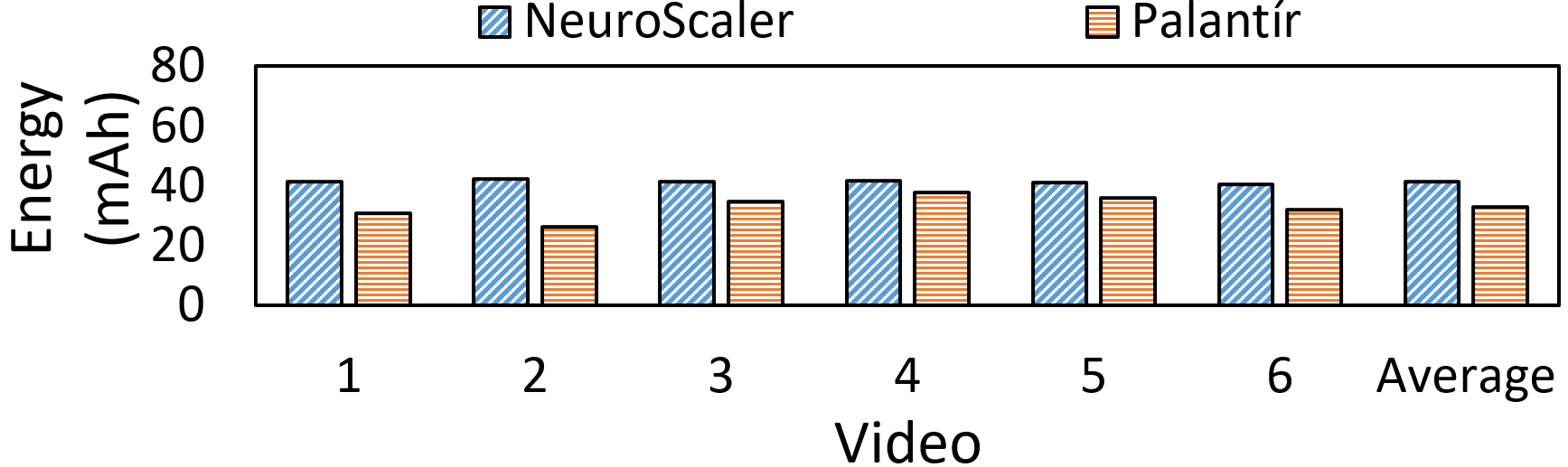}\label{fig:energy-effectiveness-2}}
\subfigure[\normalsize Setting: 3 anchor frames / 45 anchor patches.]{\includegraphics[width=0.9\linewidth]{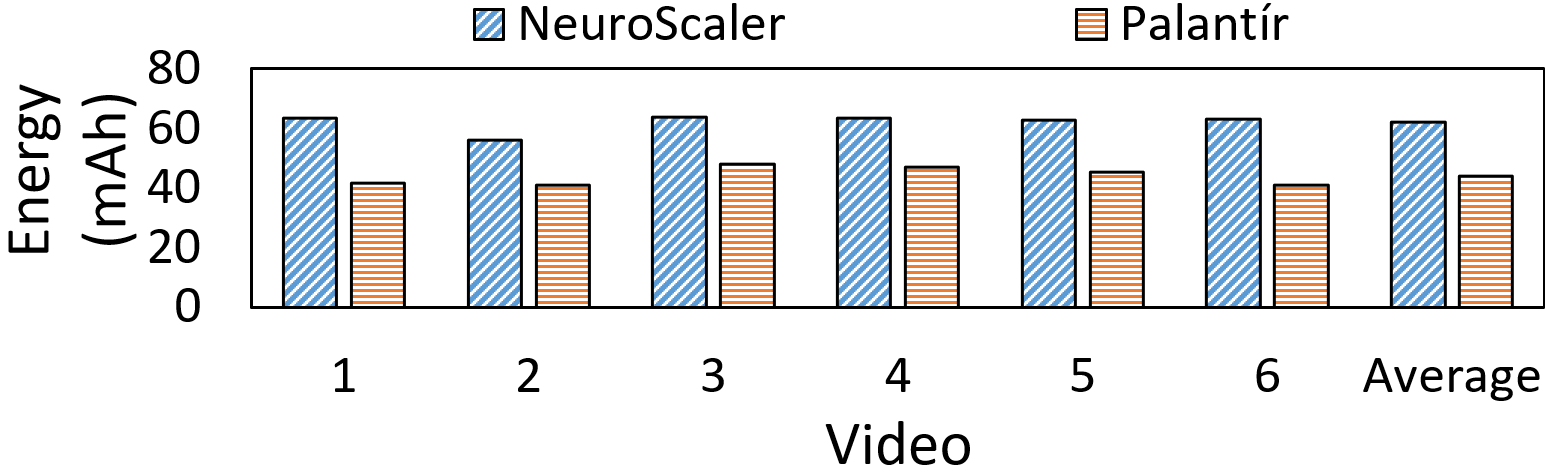}\label{fig:energy-effectiveness-3}}
\caption{A comparison of energy overhead.}
\label{fig:energy-effectiveness}

\end{figure}

\begin{figure}[t]
\setlength{\abovecaptionskip}{0pt}
\includegraphics[width=0.9\linewidth]{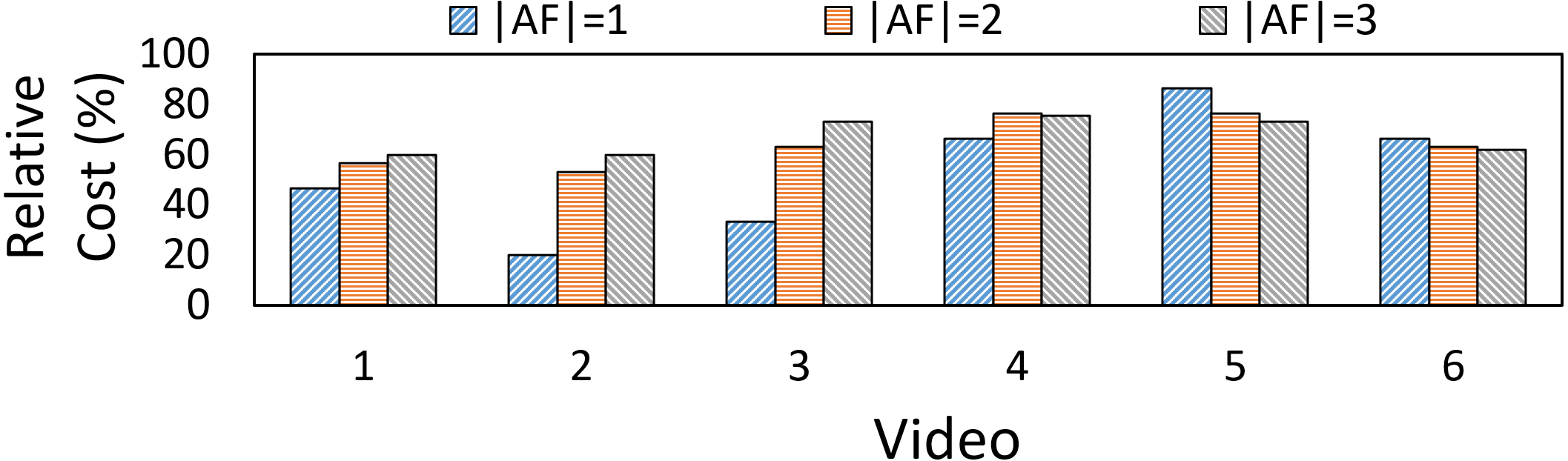}
\caption{The ratio of the monetary cost of \system to that of the NeuroScaler baseline.}
\label{fig:monetary-cost}
\end{figure}

For each anchor frame set $AF$ selected by the NeuroScaler baseline, we find the minimal anchor patch set $AP$ which is selected by \system and achieves an equivalent or higher PSNR than $AF$.
We compare the energy overhead under $AF$ with that under $AP$.
As shown in Fig.~\ref{fig:energy-effectiveness}, \system reduces the energy overhead over all cases.
The reduction ratio is 38.1\% at most and 22.4\% on average.

\para{Monetary cost reduction.}
We now present how \system reduces the monetary cost when running the SR decoder on cloud servers.
We use the same method as measuring energy efficiency to find the corresponding $AF$ for every $AP$.
The monetary cost is estimated to be linear to the DNN computation complexity under the cache profile ($AF$ or $AP$), and the keras-flops package~\cite{keras-flops} is used to measure the computation complexity.
The ratio of the monetary cost incurred by \system to that incurred by the NeuroScaler baseline is presented in Fig.~\ref{fig:monetary-cost}.
Compared to NeuroScaler, \system reduces the monetary cost by 80.1\% at most and 38.4\% on average.

\subsection{Scheduling Latency}
\label{sec:evaluation-latency}
\begin{figure}[tbp!]
\setlength{\abovecaptionskip}{0pt}
\includegraphics[width=0.8\linewidth]{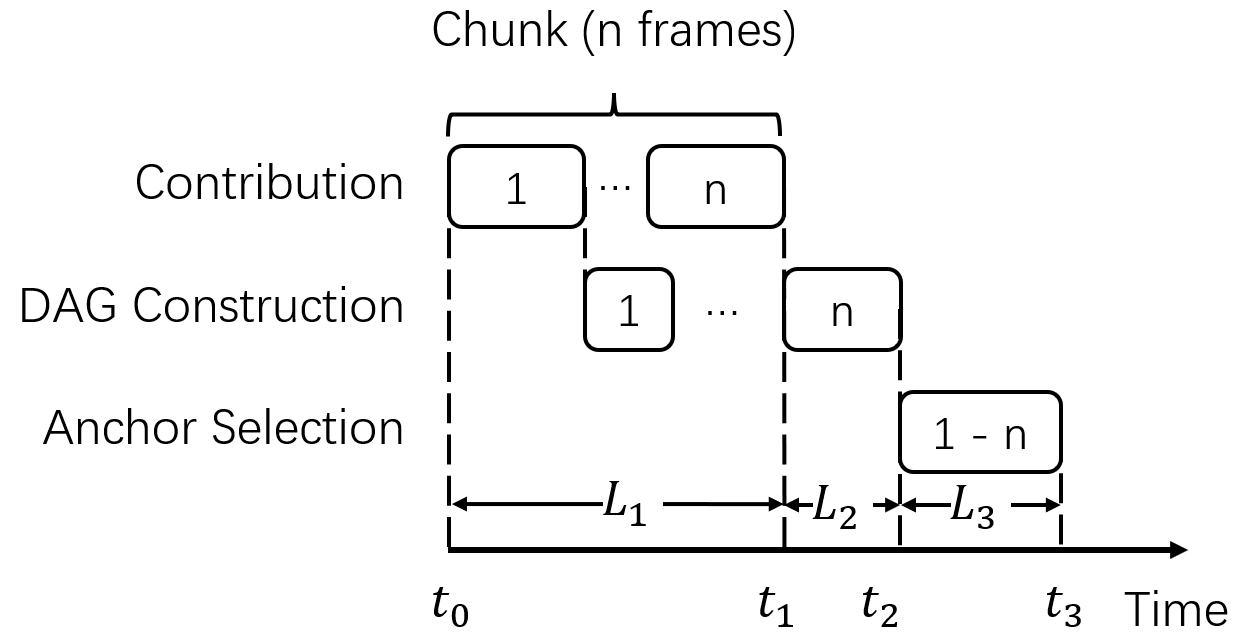}
\caption{The timeline of \system.
The numbers within the blocks represent the frame indices.}
\label{fig:latency-setting}
\end{figure}

End-to-end (E2E) latency is an important metric in live streaming~\cite{youtube-latency, GB/T-28181-2022, aws-latency, remote-drone}.
To ensure that the live streaming latency can be lower than the GoP, modern streaming standards such as CMAF~\cite{CMAF} allow a chunk (which can be part of a GoP) to be immediately packaged (\ie chunked packaging~\cite{latency-survey}) and delivered (\ie chunked delivery~\cite{latency-survey}) when ready.
Here we denote the chunk length as $n$ frames and assume the scheduling interval of \system to be equal to $n$ for simplicity.
As shown in Fig.~\ref{fig:latency-setting}, the streamer contributes new video frames at a constant rate.
Every new chunk of $n$ frames is contributed per time duration of $L_1=\frac{n}{frame\_rate}$.
In traditional streaming pipeline without neural enhancement, the new chunk can be immediately packaged and delivered at $t_1$.
However, two additional latency sources are presented in \system, \ie the DAG construction latency $L_2$ and the DAG-based anchor selection $L_3$.
We evaluate whether $L_2+L_3$ is small enough to well support latency-sensitive UHD applications.

\begin{figure}[tbp!]
\setlength{\abovecaptionskip}{0pt}
\includegraphics[width=0.7\linewidth]{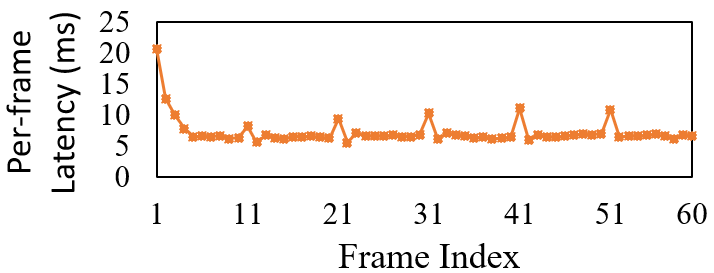}
\caption{Per-frame latency of decoding for DAG construction during a GoP.}
\label{fig:graph-construction-time}
\end{figure}

We first examine the value of $L_2$.
Note that we can directly feed a newly contributed frame to the decoder (working in the first mode introduced in Sec.~\ref{sec:evaluation-setup}) for DAG construction, so $L_2$ should be equal to the processing latency of the last frame in the scheduling interval if the decoder runs above 30fps.
As shown in Fig.~\ref{fig:graph-construction-time},  the measured per-frame decoding latency for DAG construction is indeed always below 33 ms, so we estimate $L_2$ to be the average of the measured per-frame decoding latencies in Fig.~\ref{fig:graph-construction-time}, \ie 7.2ms.

\begin{figure}[tbp!]
\setlength{\abovecaptionskip}{0pt}
\includegraphics[width=\linewidth]{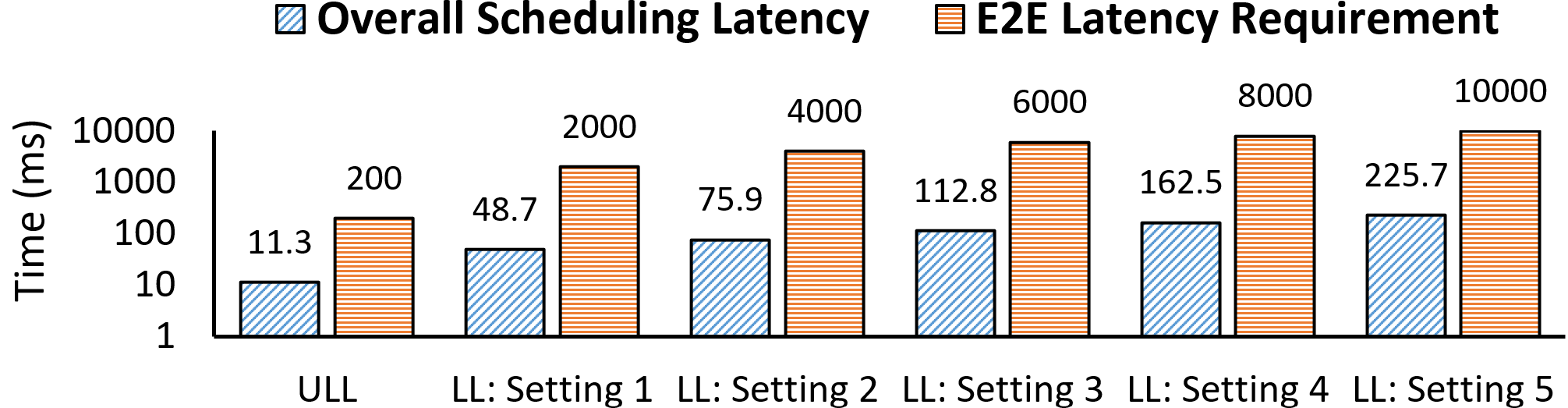}
\caption{The relationship between the overall scheduling latency and the E2E latency requirement.}
\label{fig:scheduling-latency}
\end{figure}

The DAG-based anchor selection latency $L_3$ depends heavily on the scheduling interval and the number of selected anchor patches per scheduling interval.
For ULL UHD live streaming applications~\cite{remote-drone} requiring an E2E latency below 200ms,  we set the scheduling interval to be 66.67ms (\ie 2 frames in the 30-fps evaluation videos).
As for LL live streaming applications~\cite{youtube-latency, GB/T-28181-2022, aws-latency} whose E2E latency requirements range from 2 seconds to 10 seconds, we consider five different settings (with the latency requirement being 2s, 4s, 6s, 8s, and 10s, respectively) and set the scheduling interval to be one-fifth of the latency requirement under each setting.
As illustrated in Sec.~\ref{sec:evaluation-effectiveness}, using only 5\% of all the patches as anchor patches can lead to large quality gains, so we set the ratio of searched anchor patches to be 5\% for latency measurement.
Under the above settings, the relationship between the overall scheduling latency $L_2+L_3$ (with $L_2$ fixed to 7.2ms) and the E2E latency requirement is measured and plotted in Fig.~\ref{fig:scheduling-latency}.
In all the settings of LL UHD live streaming, the overall streaming latency is less than 2.5\% of the E2E latency requirement.
As for the case of ULL live streaming, the scheduling latency is 11.3ms and accounts for about 5.7\% of the E2E latency requirement.

\subsection{Ablation Study}
\label{sec:evaluation-ablation}

\begin{figure}[t]
\setlength{\abovecaptionskip}{0pt}
\subfigure[ \normalsize Analysis on the DAG's \ attributes.]{\includegraphics[width=0.46\linewidth]{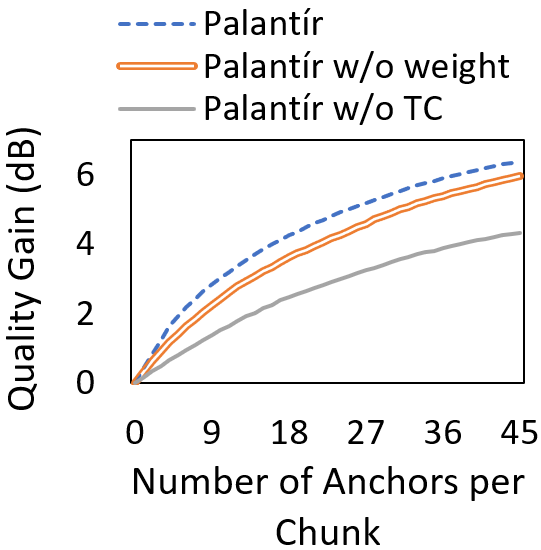}\label{fig:ablation-DAG}}
\subfigure[\normalsize Analysis on the \ parallelism mechanisms.]{\includegraphics[width=0.46\linewidth]{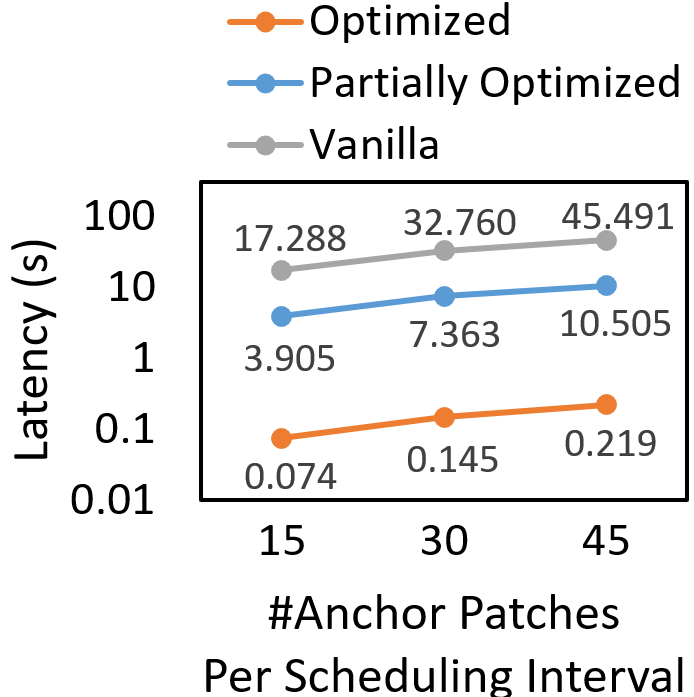}\label{fig:ablation-parallelism}}
\caption{Ablation study.}
\label{fig:ablation}
\end{figure}

\para{SR error DAG.}
The key to selecting a beneficial anchor patch set is our DAG-based modeling.
We use a theoretical analysis to determine the values of the static \textit{weight} attributes of the edges and the static \textit{TC} attributes of patch nodes (see Sec.~\ref{sec:method-estimation-analysis} and~\ref{sec:method-estimation-graph}) .
To quantify the importance of setting appropriate values, we evaluate with the \texttt{makeup review} video and compare \system with two variants.
In the first variant (\system w/o \textit{weight}), the only predecessor node of the patch node $P_n^{i,j}$ (located at the $i$-th row and $j$-th column of the patch grid of the $n$-th frame) is $P_{n-1}^{i,j}$ and the weight of the edge connecting them equals $1$.
However, the \textit{TC} attributes in the first variant are kept the same as in \system.
Note that the first variant resembles the NeuroSclaer baseline when the patch size equals the frame resolution.
In the second variant (\system w/o \textit{TC}), the \textit{weight} attributes are kept the same as in \system, but the \textit{TC} attributes of all nodes are set to $1$.
As shown in Fig.~\ref{fig:ablation-DAG}, \system consistently outperforms the two variants.

\para{Parallel Searching.}
We have introduced intra-set and inter-set parallelism (see Sec.~\ref{sec:method-parallelism}) to speed up the quality estimation process in our greedy searching algorithm.
We measure the DAG-based anchor selection latency (\ie $L_3$) under three different settings:
(1) the vanilla \system - nodes in the original DAG are processed serially for estimation;
(2) the partially optimized \system setting - only the intra-set parallelism is enabled;
(3) the optimized \system - both the two parallelism mechanisms are enabled.
The relationship between the number of anchor patches per scheduling interval and the DAG-based selection latency $L_3$ is  shown in Fig.~\ref{fig:ablation-parallelism}.
The optimized \system consistently speeds up selection by above 200 times than the vanilla \system.

%% file: future-work.tex
\section{Limitations and future work}
\label{sec:future-work}

\para{Exploring a smaller patch size.}
A natural method to further improve the efficiency of neural enhancement is to use a smaller patch size.
However, this method leads to a larger DAG and increases the latency of anchor selection.
Potential remedies may be pruning the constructed DAG.

%% file: relatedwork.tex
\section{Related Work}
\label{sec:related-work}

\para{Model Compression.}
Model compression has been utilized in many video super-resolution systems such as OmniLive~\cite{omnilive} and Microsoft Edge VSR~\cite{MSEDGE-vSR}.
Considering the heavy energy overhead of SR DNNs, it is reasonable to integrate both model compression and resuing-based SR to build a practical system.

%% file: conclusion.tex
\section{Conclusion}
\label{sec:conclusion}

In this work, we propose \system, the first neural-enhanced UHD live streaming system with fine-grained patch-level scheduling.
\system seeks to improve efficiency via reasonable scheduling while minimizing the scheduling latency to better support live streaming.
Based on our pioneering and theoretical analysis, \system adopts DAG-based quality estimation to select a beneficial anchor patch set with low computation cost.
The per-frame computation sub-procedure of the estimation method is further refactored to facilitate parallelization and acceleration and significantly decrease the scheduling latency.
The evaluation findings indicate that \system effectively optimizes the efficiency of neural enhancement and fits the latency requirement of UHD live streaming.

%% file: appendix-v3.tex
\section{Appendix}
\label{Appendix}

\subsection{Energy Measurement}
\label{sec:appendix-energy}

Here we explain how we measure the energy overhead under a given anchor set.
We use the Android Debug Bridge (adb) over Wi-Fi~\cite{adb} to execute the decoder in the smartphone's shell.
We do not use adb over USB as connecting the smartphone to an external computer via USB automatically charges the smartphone battery and affects the measured current value.
Our setting adheres mostly to the guidelines in~\cite{EdgeEnergy} for reproducibility, except that the Wi-Fi module is turned on for adb.
To remove the impact of the display screen, native daemons, and Wi-Fi interfaces in our measurements, we first record the average current $C_1$ before the decoder is executed and then record the average current $C_2$ during the execution of the decoder.
We also record the duration of decoding, $T$, and compute the energy overhead of neural-enhanced decoding as $(C_2-C_1)\times T$.